\documentclass[prx,twocolumn,showpacs,showkeys,superscriptaddress,preprintnumbers,floatfix,amsmath,amssymb,longbibliography]{revtex4-1}

\usepackage{graphicx}
\usepackage{dcolumn}
\usepackage{bm}
\usepackage{multirow}
\usepackage{array}

\usepackage[usenames,dvipsnames]{xcolor}
\definecolor{nblue}{rgb}{0.0, 0.0, 1.0}
\definecolor{magenta}{rgb}{0.79, 0.08, 0.48}
\usepackage[colorlinks,linkcolor=nblue,urlcolor=magenta,citecolor=magenta,plainpages=false,pdfpagelabels,breaklinks]{hyperref}
\usepackage{changes}

\newcommand{\beq}{\begin{equation}}
\newcommand{\eeq}{\end{equation}}
\newcommand{\bea}{\begin{eqnarray}}
\newcommand{\eea}{\end{eqnarray}}

\begin{document}

\title{Weak antilocalization in the transition metal telluride Ta$_2$Pd$_3$Te$_5$}

\author{Wen-He Jiao}
\email[]{whjiao@zjut.edu.cn}
\affiliation{Key Laboratory of Quantum Precision Measurement of Zhejiang Province, School of Physics, Zhejiang University of Technology, Hangzhou 310023, China}

\author{Hang-Qiang Qiu}
\affiliation{Department of Applied Physics, Zhejiang University of Science and Technology, Hangzhou 310023, China}

\author{Wuzhang Yang}
\affiliation{Department of Physics, School of Science, Westlake University, Hangzhou 310024, China}
\affiliation{Institute of Natural Sciences, Westlake Institute for Advanced Study, Hangzhou 310024, China}

\author{Jin-Ke Bao}
\affiliation{Department of Physics, Materials Genome Institute and International Center for Quantum and Molecular Structures, Shanghai University, Shanghai 200444, China}

\author{Shaozhu Xiao}
\affiliation{Ningbo Institute of Materials Technology and Engineering, Chinese Academy of Sciences, Ningbo 315201, China}

\author{Yi Liu}
\affiliation{Key Laboratory of Quantum Precision Measurement of Zhejiang Province, School of Physics, Zhejiang University of Technology, Hangzhou 310023, China}

\author{Yuke Li}
\affiliation{School of Physics and Hangzhou Key Laboratory of Quantum Matters, Hangzhou Normal University, Hangzhou
311121, China}

\author{Guang-Han Cao}
\affiliation{School of Physics, Interdisciplinary Center for Quantum Information, and State Key Laboratory of Silicon and Advanced Semiconductor Materials, Zhejiang University, Hangzhou 310058, China}

\author{Xiaofeng Xu}
\affiliation{Key Laboratory of Quantum Precision Measurement of Zhejiang Province, School of Physics, Zhejiang University of Technology, Hangzhou 310023, China}

\author{Zhi Ren}
\affiliation{Department of Physics, School of Science, Westlake University, Hangzhou 310024, China}
\affiliation{Institute of Natural Sciences, Westlake Institute for Advanced Study, Hangzhou 310024, China}

\author{Peng Zhang}
\affiliation{School of Physics and National Laboratory of Solid State Microstructures, Nanjing University, Nanjing 210093, China}

\date{\today}

\begin{abstract}
We report transport studies on the layered van der Waals topological crystalline insulator Ta$_2$Pd$_3$Te$_5$. The temperature-dependent resistance at high temperature is dominated by a bulk insulating gap and tend to saturate at low temperatures. Low temperature magnetotransport shows that Ta$_2$Pd$_3$Te$_5$ exhibits weak antilocatization (WAL) effect in both perpendicular orientation and parallel orientation, suggesting an contribution of the WAL effect from both topological edge states and bulk states. By measuring the anisotropic magnetoconductance and then subtracting the contribution of bulk states, the WAL effect associated with topological edge states can be revealed and analyzed quantitatively based on the two-dimensional Hikami-Larkin-Nagaoka model. Our results have important implications in understanding the WAL phenomena in Ta$_2$Pd$_3$Te$_5$.
\end{abstract}

\maketitle

\section{Introduction}

In the quantum diffusive regime, the quantum interference between time-reversed scattering loops can give rise to a quantum correction to the conductivity \cite{1980-PTP,Shen-2011prl}. The constructive superposition of the scattering amplitudes gives rise to a negative correction to
the conductivity, known as quantum weak locatization (WL), while in electronic systems with sufficiently strong spin-orbit coupling (SOC), spin
procession along closed electron trajectories would lead to reduced backscattering, and thus a positive correction, which is so-called quantum weak antilocatization (WAL). Both WL and WAL can be suppressed under external magnetic fields as a result of time-reversal symmetry breaking. The diminished destructive interference effects result in a negative magnetoresistance for WL \cite{WL} and a positive magnetoresistance for WAL \cite{WAL}. As the probability of two interfering scattering pathways is larger in lower dimensions, WL and WAL are generally more pronounced in
low-dimensional systems \cite{PRB-Chu}.

While in conventional metallic compounds or semiconductors with a parabolic band dispersion, WAL results
from strong SOC, it may also occur in systems with a Dirac-like band dispersion \cite{PRL-Ando}. In such systems as quantum spin Hall insulators
(QSHIs), topological insulators (TIs) and topological crystalline insulators (TCIs), electrons' spin in topological edge/surface states gives rise to a nontrivial $\pi$ Berry phase of carriers at the Fermi energy, due to which carrier backscattering can be suppressed. The WAL effects stemming from surface states of TIs and TCLs have been frequently observed and extensively studied in the past decades \cite{PRL-Ong,PRL-ZFC,NL-HJ}.

Recently, the van der Waals (vdW) monolayers Ta$_2$Pd$_3$Te$_5$ were predicted to be potential candidates for hosting QSH states
based on first-principles calculations \cite{PRB-WZJ}. The open-boundary calculations confirm the existence of nontrivial helical edge
states therein. Soon after, Wang $et$ $al$. reported the direct observation of QSH states, namely the topological edge states by scanning tunneling microscopy/spectroscopy (STM/STS) measurements in the vdW material Ta$_2$Pd$_3$Te$_5$ \cite{PRB-Feng}. Due to the double-band inversion,
the monolayer was further predicted to be a two-dimensional quadrupole topological insulator with second-order topology \cite{NPJ-Ta235}. Although the bulk Ta$_2$Pd$_3$Te$_5$ was proposed to be TCIs with trivial symmetry indicators as the Chern numbers in the $k_y$ = 0 plane are computed to be nonzero \cite{PRB-WZJ}, a dip-like feature of the magnetoresistance at low magnetic fields is attributed to the WAL effect, which indicates the presence of quantum transport characteristics of TCIs in bulk Ta$_2$Pd$_3$Te$_5$ \cite{APL-Tian}. More recently, the Tomonoga-Luttinger liquid inhabiting the edge states was observed in Ta$_2$Pd$_3$Te$_5$, demonstrating Ta$_2$Pd$_3$Te$_5$ has the two-dimensional second-order topology with correlated edge states \cite{Arxiv-Shen}. The appearance of superconductivity induced by Ti- and W-doping, and the metallization and emergence of superconductivity via pressure manipulation in bulk Ta$_2$Pd$_3$Te$_5$, further indicates either monolayer or bulk form serves as a good platform for studying the novel quantum states and the interactions in low dimensions \cite{JPSJ-Ta235,Arxiv-Ta235}.

In this work, we report the temperature-dependent resistivity, Hall effect, anisotropic magnetoresistance studies on Ta$_2$Pd$_3$Te$_5$ single crystals, which are grown by the method of chemical vapor transport. The resistivity shows a semiconductor behavior from 300 K to 100 K, and crosses
over to a three-dimensional variable-range hopping (VRH) behavior until 18 K. At low temperatures, a large bulk resistivity exceeding 2 $\Omega$ cm
can be found and tends to saturate, indicating a metallic contribution. In this temperature range, the Hall coefficient also shows a slight decrease. In both perpendicular and parallel field configurations, low-temperature magnetoconductance exhibits WAL effect, suggesting an contribution of the WAL effect from both topological edge states and bulk states. By measuring the anisotropic magnetoconductance, the two-dimensional WAL effect, possibly associated with topological edge states, is revealed and analyzed quantitatively based on the two-dimensional Hikami-Larkin-Nagaoka (HLN) model.

\section{Experimental methods}

\begin{figure}
\begin{center}
\includegraphics[width=0.5\textwidth]{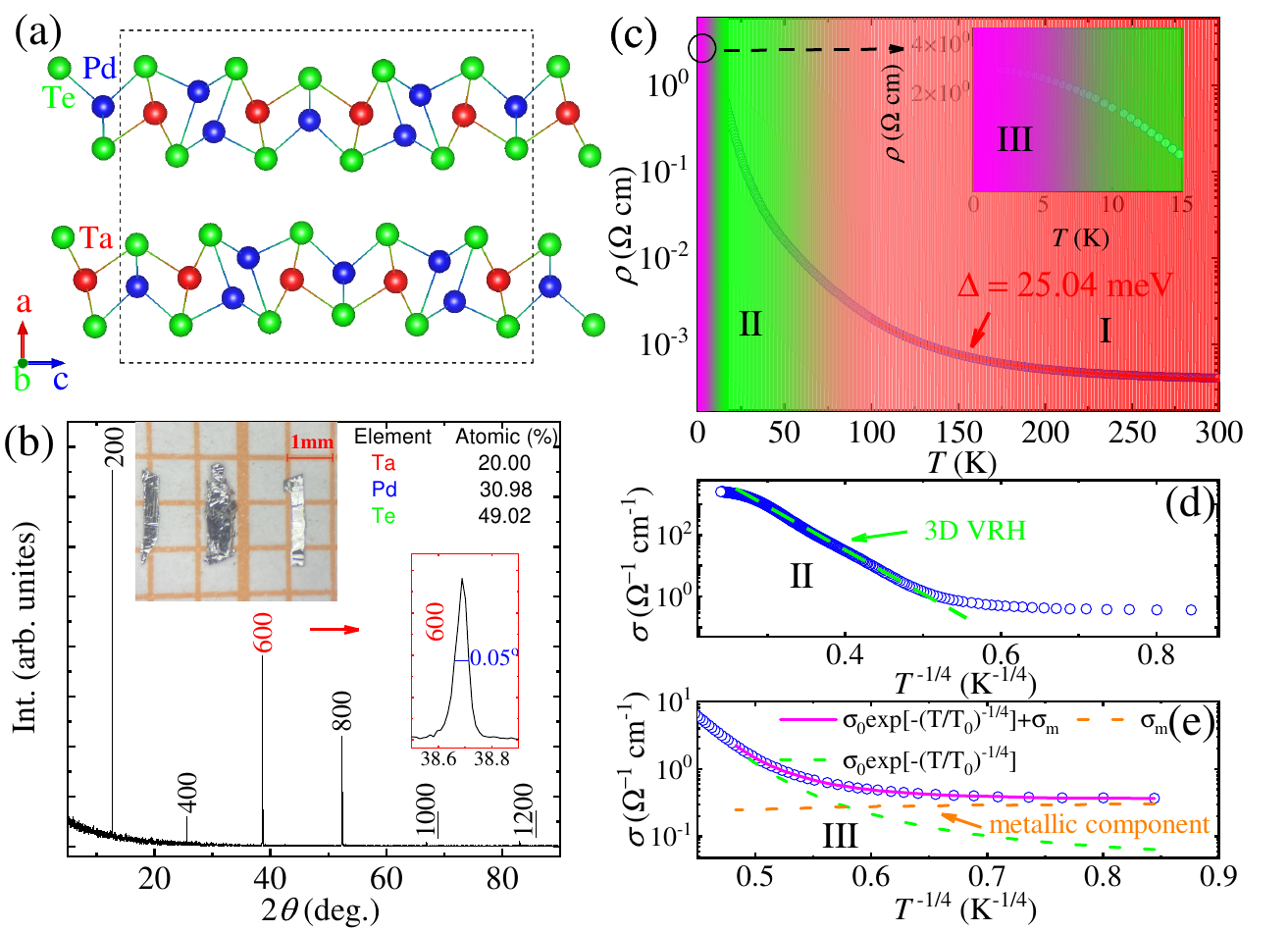}
\caption{\label{fig1}(Color online) (a) The crystallographic structure of Ta$_2$Pd$_3$Te$_5$ viewed along the $b$ axis. (b) Single-crystal x-ray diffraction pattern at room temperature. The right inset enlarges the third reflection in the x-ray diffraction pattern. The left inset is a photograph of the as-grown Ta$_2$Pd$_3$Te$_5$ crystals on a millimeter paper. (c) Temperature dependence of the electrical resistivity $\rho$ with the current applied along the $b$ axis shown in a logarithmic plot. The red line is the fit to the Arrhenius formula above 100 K as region I. The insets enlarge the low-temperature region, from which one can see the tendency of the saturation. (d) Plot of the conductivity $\sigma$ versus $T^{-1/4}$. The dashed green line is the fit to the three-dimensional (3D) VRH formula for region II. (e) Plot of the conductivity $\sigma$ versus $T^{-1/4}$ at low temperature. The magenta line is the fitting result for region III with the formula taking into account a metallic conduction (dashed orange line) on the basis of 3D VRH behavior (dashed green line).}
\end{center}
\end{figure}

High-quality Ta$_2$Pd$_3$Te$_5$ single crystals were obtained by chemical vapor transport.
Powders of the elements Ta(99.97\%), Pd(99.995\%), and Te(99.999\%) were weighed in a stoichiometric amount, and were thoroughly ground by hand in an agate mortar. About 5\%wt iodine as transport agent were then added to the mixture. The final mixture was loaded, and sealed into an evacuated quartz ampule. All the procedures handling the reagents were done in a glove box filled with pure argon gas (O$_2$ and H$_2$O $<$ 0.1 ppm). The ampoule was slowly heated to 1123 K and 1073 K in the respective heating zone for 24 hours.
The polarity of the temperature gradient was reversed after one week and held for another week. The air-stable crystals with shiny gray-black flattened needle-like shape were finally obtained in cooler side of the ampule. The schematic diagram showing the transport process is shown in Fig. S1 of the supplemental material (SM) \cite{SM}. The size of the as-grown crystals are $\sim$ 2$\times$1$\times$0.2 mm$^3$ [the photographic image shown in the left inset of Fig.~\ref{fig1}(b)], and can be easily exfoliated to a thin layer by a razor blade.

The X-ray diffraction (XRD) data acquisition from the crystal was performed with a monochromatic Cu K$_{\alpha 1}$ radiation using a PANalytical x-ray diffractometer radiation by a conventional $\theta$-2$\theta$ scan (Model EMPYREAN). The crystal was mounted on a sample holder, which is then followed by the data collection at room temperature. The crystal structure is plotted using the software VESTA \cite{VESTA}. Energy-dispersive x-ray spectroscopy (EDS) was carried out by using the Hitachi S-3400 instrument to get the chemical composition of the as-grown crystals. The EDS analysis was performed on the fresh surface of three selected crystals, which gives the average composition of Ta$_{2.0}$Pd$_{3.1}$Te$_{4.9}$. The detailed EDS results are tabulated in Table S1 of the SM, as seen from which the slight variations of the chemical composition on different positions in either crystal indicate the presence of impurity and/or defect centers.

The temperature-dependent resistivity, Hall effect, and temperature-/angel-dependent magnetotransport measurements,  were performed in the Physical Property Measurement System (PPMS-9, Quantum Design) with the standard four-probe method. The Hall-effect measurement was performed by reversing the field direction. In this way, a small field-symmetric component due to misaligned electrodes can be subtracted from the Hall data. An angle-resolved photoemission spectroscope (ARPES) with laser (photon energy h$\nu$ = 6.994 eV) and a base pressure better than 5 $\times$ 10$^{-11}$ mbar was utilized to measure the band structure of the samples, and measurements were performed on a fresh (100) surface of a separate crystal from the same batch. The ARPES measurement was carried out at 10 K.

\section{Results and Discussion}

Ta$_2$Pd$_3$Te$_5$ crystallizes in the space group $Pnma$ with the lattice parameters: $a$ = 13.9531(6) {\AA}, $b$ = 3.7038(2) {\AA}, $c$ = 18.5991(8) {\AA} \cite{PRB-WZJ}. Each unit cell contains two Ta$_2$Pd$_3$Te$_5$ monolayers stacked along the $a$ axis via weak vdW interactions. Each monolayer
is composed of three atomic layers. Te atoms form the top and bottom layers, between which the middle layers are formed by Ta and Pd atoms. The arrangement of one distorted octahedral Pd$_2$Te$_2$ chains, two distorted trigonal prismatic TaTe and one trigonal prismatic PdTe chains,
that run parallel to the $b$ axis, constitutes the layered slab. The crystallographic structure viewed along the $b$ axis is shown in Fig.~\ref{fig1}(a). The crystal morphology, as shown in the left inset of Fig.~\ref{fig1}(b), reflects its structural characteristics.
Figure ~\ref{fig1}(b) shows the XRD pattern of the crystals with the layered facet lying on the sample holder at room temperature. A set of observed diffraction peaks can be well indexed with the (\underline{2$l$}00) peaks, which suggests the $a$ axis being perfectly perpendicular to the layered facet of the crystal. The full width at half-maximum of the diffraction is as small as 0.05$^\circ$ for the (600) peak. From the peak position, we can calculate the average inter-planar spacing to be 6.9737(2) {\AA}, very close to half of the reported lattice parameter $a$ ($a$/2 = 6.9765(5) {\AA}).

The temperature dependence of the electrical resistance $\rho$($T$) along the $b$ axis is displayed in Fig.~\ref{fig1}(c). The semiconductor behavior above 100 K, labeled as region I, can be well fitted with the Arrhenius model $\rho$ $\sim$ exp($\Delta$/k$_\textrm{B}T$), where $\Delta$ is the activation energy and k$_\textrm{B}$ is
the Boltzmann constant. The red line in Fig.~\ref{fig1}(c) shows the fitting results with the fitted $\Delta$ $\sim$ 25.04 meV, much larger than that
in Ref. \cite{PRB-Feng}. This difference may arise from the different quality of single crystals grown by different methods. Below the temperature
range of the activated behavior, $\rho$($T$) in region II appears to be better described by the 3D VRH behavior, $\rho$ $\sim$ exp[($T$/$T_\textrm{0}$)$^{-1/4}$],
where $T_\textrm{0}$ is a constant that depends on the density of states at the Fermi level \cite{EPDS}. With the temperature decreasing further in region III, the resistivity shows a tendency of saturation. As shown in Fig.~\ref{fig1}(e), we employ a two-channels model to understand the conductivity ($\sigma$=1/$\rho$) in this region, i.e., a
parallel circuit of an insulating component characterized by the 3D VRH behavior and a metallic component [$\sigma_\textmd{m}$=($\rho_{\textmd{is}}$+$AT$)$^{-1}$], where $\rho_{\textmd{is}}$ is a low-temperature residual resistance due to impurity
scattering and $A$ is an electron-phonon scattering constant \cite{PRB-Egger}. The similar metallic channel was associated with the edge/surface
states of Ta$_2$Pd$_3$Te$_5$ \cite{APL-Tian}, as a common approach for interpreting the resistivity saturation at low temperature in the topologically nontrivial materials \cite{NP-2014,Npj-Sun}. However, a degenerate bulk impurity band could also contribute to the low-temperature saturated resistivity \cite{PRB-Ren}, which follows the same temperature dependence as 1/$\sigma_\textmd{m}$ \cite{JMMM-2000}. The fitting parameters extracted here, along with those in Ref. \cite{APL-Tian}, are summarized in Table S2 of the SM as a comparison.

\begin{figure}
\includegraphics[width=0.48\textwidth]{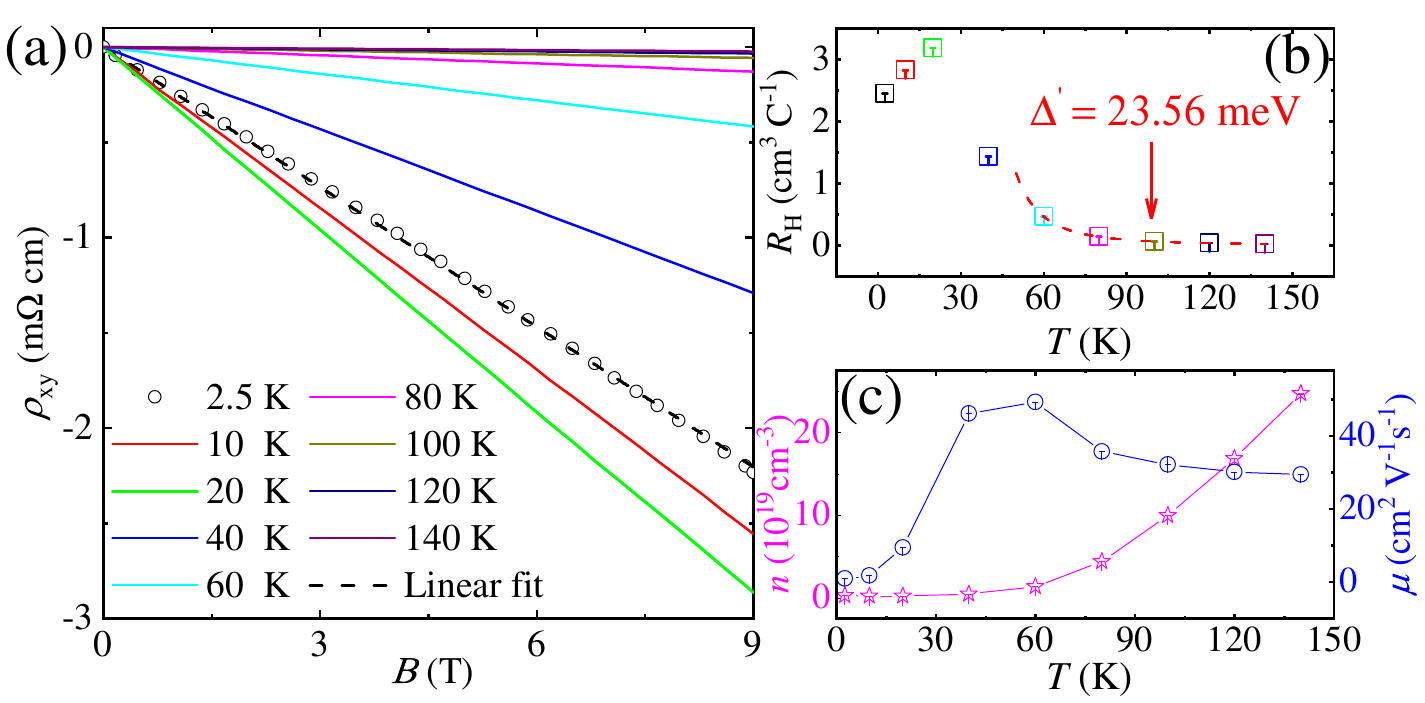}
\caption{\label{Hall} (Color online) (a) The Hall resistivity of Ta$_2$Pd$_3$Te$_5$ at selected temperatures below 140 K. The dashed line is a linear fit. (b) The temperature-dependent Hall coefficient derived from the linear fits. (c) Temperature dependence of the extracted carrier density (left axis) and mobility (right axis).}
\end{figure}

Figure~\ref{Hall}(a) displays the magnetic field dependence of the Hall resistivity $\rho_{\textmd{xy}}$ at selected temperature below 140 K. The negative $\rho_{\textmd{xy}}$ linearly decreases with magnetic field, suggesting the dominance of electron-type charge carriers. The Hall coefficient $R_\textmd{H}$($T$) obtained by linear fitting is plotted in Fig.~\ref{Hall}(b). At high temperature ($\geq$ 60 K), the behavior of $R_\textmd{H}$($T$) is thermally activated, and the data can be fitted with the Arrhenius model. The fitted activation energy $\Delta^{\textmd{'}}$ = 23.56 meV is very close to that we obtain from the resistivity analysis above, though the temperature ranges for the two fittings are slightly different. With the temperature decreasing in region III, a slight decrease of $R_\textmd{H}$ from 3.19 cm$^3$/C at 20 K to 2.45 cm$^3$/C at 2.5 K can be observed, corresponding to the carrier density $n$ of 1.96 $\times$ 10$^{18}$ cm$^{-3}$ at 20 K and 2.55 $\times$ 10$^{18}$ cm$^{-3}$ at 2.5 K in a one-band model. The shift of the chemical potential as temperature varies may cause the temperature dependence of $n$, as shown in Fig.~\ref{Hall}(c). The blue symbols in Fig.~\ref{Hall}(c) summarize the values of Hall mobility, $\mu$
= $R_\textmd{H}$/$\rho$, as a function of temperature. The extremely low mobility at low temperatures excludes the possible origin of charge carriers from the edge states but suggests the origin from the impurity band.

\begin{figure*}
\includegraphics[width=1.02\textwidth]{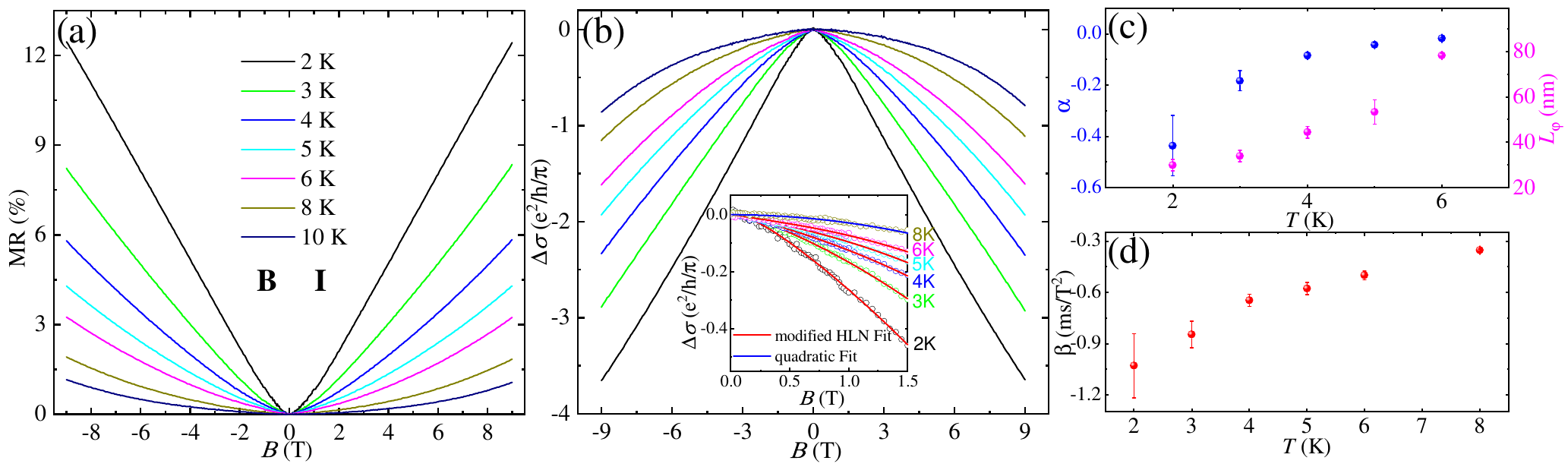}
\begin{center}
\caption{\label{MRT} (Color online) (a) The magnetoresistance measured at different temperatures indicated with applied magnetic field perpendicular to the current. (b) The magnetoconductance as a function of magnetic field. The inset displays the fitting curves using the modified Hikami-Larkin-Nagaoka (HLN) model. (c) Temperature dependence of the fitting parameters $\alpha$ (left axis) and $l_\varphi$ (right axis). (d) Temperature depencence of the fitting parameter $\beta$.}
\end{center}
\end{figure*}

In Fig.~\ref{MRT}(a), we present the magnetoresistance (MR), defined as MR(\%) = $\frac{\rho(B)-\rho(0 \textrm{T})}{\rho(0 \textrm{T})}$ $\times$ 100\%, of the Ta$_2$Pd$_3$Te$_5$ single crystal measured with the applied current along the $b$-axis and the applied magnetic field along the $a$-axis ($\emph{\textbf{B}}$$\perp$$\emph{\textbf{I}}$) up to 9 T at varied temperatures below 10 K. All the curves display the positive value without saturation up to 9 T, and the negative MR at low fields above 5 K, as reported in Ref. \cite{APL-Tian}, doesn't appear in our data. The small negative MR has been attributed to the WL effect, which usually appears in disordered electronic
systems. The different sample quality of the crystals may be the possible reason for the disappearance of the small negative MR here as the level of deficiencies and disorders can play a key role in giving rise to the negative MR \cite{WL}. The MR at 2 K and 9 T reaches 12.4\%, and gradually decreases with the temperature increasing. The behavior of MR at low fields shows a dip-like feature at low temperatures, indicating the presence of a WAL effect \cite{PRL-Lu}, and gradually evolves into a parabolic one above 6 K. The WAL effect is more pronounced by the field dependence of the sheet magnetoconductance $\Delta$$\sigma$ = ($L$/$W$)[1/$R$($\emph{\textbf{B}}$) $-$ 1/$R$(0)], where $L$ and $W$ are the length and width of the thin crystal measured respectively, and $R$($\emph{\textbf{B}}$) is the resistance under the applied magnetic field along the $a$-axis, as shown in Fig.~\ref{MRT}(b). The WAL behavior can originate either from strong SOC in the bulk or
from spin-momentum locking in the topological systems \cite{PRL-Ando}. In the low mobility and strong SOC regime, the HLN equation is often applied to characterize the WAL effect \cite{PTP-HLN}. If one take\added{s} into account the background with parabolic conductance contribution, the modified HLN
formula can be written as \cite{APL-Tian,SR-2016}:
\begin{align}\label{eq.1}
\Delta \sigma = \alpha(\frac{e^2}{2\pi^2\hbar})[\Psi(\frac{1}{2}+\frac{B_\phi}{B})-\ln(\frac{B_\phi}{B})] + \beta B^2,
\end{align}
where $\Psi$($x$) is the digamma function, and $B_\phi$ = $\hbar$/(4$el_\varphi^2$) is the phase coherence characteristic field with $l_\varphi$
being the phase coherence length. This model can describe the data satisfactorily below 6 K and the fitting curves are shown in the inset of Fig.~\ref{MRT}(b), while the data at 8 K follows the semiclassical $B^2$ dependence resulting from the Lorentz deflection of charge carriers. The parameter $\alpha$ is a reflection of the number of independent conduction channels contributing to the interference \cite{PRB-2011shen}. For the WAL effect induced by the spin-orbit interaction in systems with parabolic dispersion and no magnetic scattering (the symplectic case), the value of $\alpha$ should be $-$1/2 \cite{PTP-1980}. For the quantum spin texture systems such as topological surface states in TLs, the value of $\alpha$ should also be $-$1/2 for a single surface channel due to the existence of nontrivial $\pi$ Berry phase \cite{PRL-ZFC,PRL-2010}. The modified HLN fitting at 2 K yields the reasonable parameters $\alpha$ = $-$0.43 $\pm$ 0.12, $l_\varphi$ = 30 $\pm$ 3 nm, $\beta$ = $-$1.03 $\pm$ 0.19 mS T$^{-2}$, the temperature dependence of which are plotted in Figs.~\ref{MRT}(c-d). However, the value of $l_\varphi$ should decrease with the temperature increasing due to the dephasing mechanism \cite{NL-2012}, contrary with what we observed here. Its possible origin is an open question and still under investigation.

\begin{figure}
\includegraphics[width=0.49\textwidth]{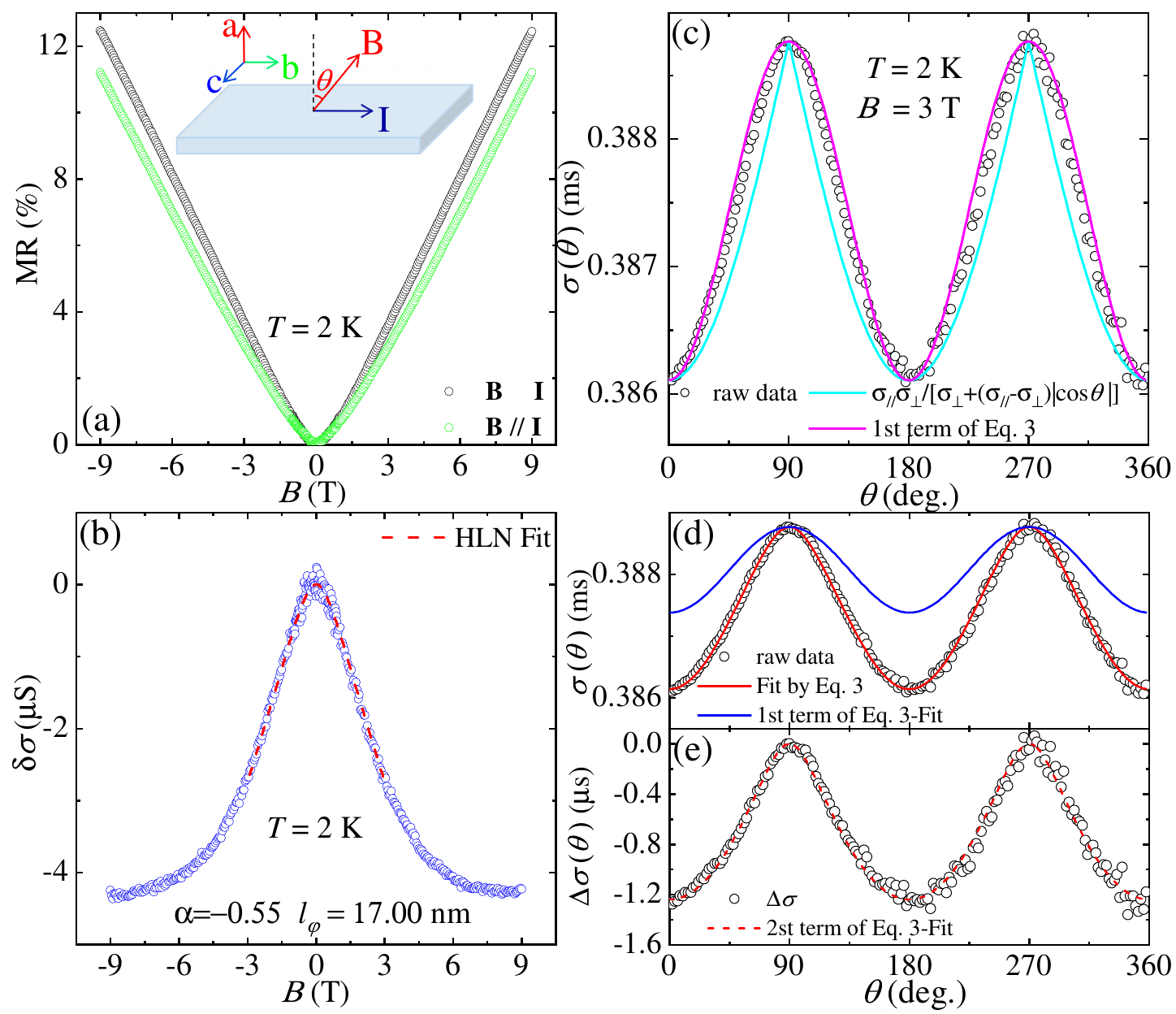}
\caption{\label{angle} (Color online) (a) The magnetoresistance measured at 2 K with applied magnetic field both perpendicular and parallel to the current. The inset shows the schematics of the experiment and the definition of the angle $\theta$. (b) The magnetoconductance difference $\Delta$$\sigma$($\theta$=90$^\circ$, $\emph{\textbf{B}}$) as a function of magnetic field at 2 K. The red dotted line indicates the fitting curve using HLN equation. (c) The angular dependence of the magnetoconductance $\sigma$ at 2 K and 3 T. The cyan solid line is the fitting curve using
$\frac{\sigma_{\parallel}\sigma_{\bot}}{\sigma_{\bot}+(\sigma_{\parallel}-\sigma_{\bot})|\cos\theta|}$ relationship. The violet solid line is the fitting curve of the first term of Eq.~\ref{eq.3} using measured $\sigma_{\parallel}$ and $\sigma_{\bot}$ values. (d) The angular dependence of the magnetoconductance $\sigma$ and the red solid line is the fitting curve using Eq.~\ref{eq.3} with $\sigma_{\bot}$ = 0.387 mS, $\alpha$ = $-$0.28 and $l_\varphi$ = 16.2 nm. The blue solid line is the fitting curve of the first term of Eq.~\ref{eq.3} using measured $\sigma_{\parallel}$ value and fitted $\sigma_{\bot}$ value of about 0.387 mS. (e) The angular dependence of the $\Delta \sigma$, which is the difference between $\sigma$($\theta$) measured and the blue curve in panel (d) (the first term of Eq.~\ref{eq.3}), at 2 K and 3 T. The red dotted line is the fitting curve using HLN expression with $\alpha$ = $-$0.28 and $l_\varphi$ = 16.2 nm.}
\end{figure}

We have also carried out the MR measurement on the same Ta$_2$Pd$_3$Te$_5$ crystal with the applied magnetic field parallel to the current direction ($\emph{\textbf{B}}$$\parallel$$\emph{\textbf{I}}$) up to 9 T at 2 K. The value of the MR for $\emph{\textbf{B}}$$\parallel$$\emph{\textbf{I}}$ is slightly smaller than that for $\emph{\textbf{B}}$$\perp$$\emph{\textbf{I}}$, as shown in Fig.~\ref{angle}(a). The similar dip-like feature displayed by the
behaviors of the temperature-dependent MR for both cases indicates that the presence of the WAL effect in both field orientations, i.e. the presence of
a three-dimensional
component of the WAL effect. Generally, MR should be isotropic if WAL is purely caused by the SOC in a three-dimensional bulk channel \cite{PRB-Chu}.
The scenario of the MR curves suggests the superposition of two contributions to the WAL effect, one from bulk states and one from topological edge states. We can reveal the WAL effect induced from topological edge states by subtracting the three-dimensional WAL contribution from the sheet magnetoconductance data for $\emph{\textbf{B}}$$\perp$$\emph{\textbf{I}}$ using $\delta$$\sigma$ = ($L$/$W$)[1/$R$(0$^\circ$,$\emph{\textbf{B}}$) $-$ 1/$R$(90$^\circ$,$\emph{\textbf{B}}$)] \cite{PRL-ZFC}. Figure~\ref{angle}(b) displays $\delta$$\sigma$ as a function of the field, the contribution ratio of which to the conductivity for $\emph{\textbf{B}}$$\perp$$\emph{\textbf{I}}$ is very small ($\sim$0.68\% at 3 T). A sharper dip is clearly observed, indicative of the WAL effect due to the topological edge states in Ta$_2$Pd$_3$Te$_5$.
We also characterize the WAL effect by employing the two-dimensional HLN equation \cite{PTP-HLN}:
\begin{align}\label{eq.2}
\delta \sigma = &\alpha(\frac{e^2}{2\pi^2\hbar})[\Psi(\frac{1}{2}+\frac{B_\phi}{B})-\ln(\frac{B_\phi}{B})],
&\end{align}
As shown by the red dashed fitting curve in Fig.~\ref{angle}(b), the magnetoconductance tip can be well fitted with this
formula. The fitting yields $\alpha$ = $-$0.55 $\pm$ 0.02 and $l_\varphi$ = 17.0 $\pm$ 0.3 nm, respectively.

If the magnetoconductance of the bulk states is anisotropic, a more accurate method can be employed to eliminate the contribution of bulk states for
quantitatively analyzing the WAL effect of topological states \cite{NL-Wang}. Figure~\ref{angle}(c) shows the symmetrized magnetoconductance $\sigma$($\theta$) as a function of the rotating angle $\theta$ measured at 2 K and 3 T. $\sigma$($\theta$) displays a 180$^\circ$ periodic angular dependence with the minima and maxima appearing at $\emph{\textbf{B}}$$\perp$$\emph{\textbf{I}}$ ($\theta$ = 0$^\circ$ or 180$^\circ$) and $\emph{\textbf{B}}$$\parallel$$\emph{\textbf{I}}$ ($\theta$ = 90$^\circ$ or 270$^\circ$), respectively. In general, the conventional anisotropic
magnetoconductance should follow as $\frac{\sigma_{\parallel}\sigma_{\bot}}{\sigma_{\bot}+(\sigma_{\parallel}-\sigma_{\bot})\cos^2\theta}$ \cite{NL-Wang,PRB-MR}, where $\sigma_{\parallel}$ and $\sigma_{\bot}$ are the conductance measured in the parallel and perpendicular field configurations, respectively. As shown by the violet solid fitting line using this equation in Fig.~\ref{angle}(c), there is a small deviation from the experimental data. For two-dimensional surface states, the magneto-transport is expected to vary with the perpendicular component of the magnetic field $B\cos$($\theta$). In this case, $\sigma$($\theta$) should have wide peaks around the perpendicular field configuration ($\theta$ = 0$^\circ$ or 180$^\circ$) and dips with cusp around the parallel field configuration ($\theta$ = 90$^\circ$ or 270$^\circ$). Then, a $\frac{\sigma_{\parallel}\sigma_{\bot}}{\sigma_{\bot}+(\sigma_{\parallel}-\sigma_{\bot})|\cos\theta|}$ relationship is expected to be followed
by $\sigma$($\theta$) \cite{NL-Wang}. The cyan solid fitting line with this formula obviously mismatches with the experimental data as well.
The QSH states are confirmed in Ta$_2$Pd$_3$Te$_5$ \cite{PRB-Feng}, and the WAL effect due to the topological edge states can be fairly described by the HLN equation as we discussed above. Following a previous approach, we employ an equation that includes both the conventional anisotropic magnetoconductance and the two-dimensional HLN-type WAL contribution to interpret the angle-dependent magnetoconductance measured here:
\begin{align}\label{eq.3}
 \sigma = &\frac{\sigma_{\parallel}\sigma_{\bot}}{\sigma_{\bot}+(\sigma_{\parallel}-\sigma_{\bot})\cos^2\theta} + \notag\\ &\alpha(\frac{e^2}{2\pi^2\hbar})[\Psi(\frac{1}{2}+\frac{B_\phi}{|B\cos\theta|})-\ln(\frac{B_\phi}{|B\cos\theta|})],
\end{align}
The first term in the above equation is the anisotropic magnetoconductance originated from the bulk states and the second term is that induced by the
topological edge states due to the Berry curvature. As there is no contribution from the second term in the parallel field configuration, we take
$\sigma_{\parallel}$ being the measured value but $\sigma_{\bot}$ being a fitting parameter in the analysis.
As shown by the red solid fitting curve in Fig.~\ref{angle}(e), the experimental data can be produced perfectly by this equation with
$\sigma_{\bot}$ = 0.387 $\pm$ 0.001 mS, $\alpha$ = $-$0.28 $\pm$ 0.14, and $l_\varphi$ = 16.2 $\pm$ 2.1 nm. The first term, originated from the bulk states, is plotted as the blue line in Fig.~\ref{angle}(d), which has wide peaks around the both perpendicular and parallel field configurations.
The difference of the angular-dependent mganetoconductance $\Delta \sigma$($\theta$) between $\sigma$($\theta$) and the blue curve in Fig.~\ref{angle}(d), along with the red dotted fitting curve from the second term of Eq.~\ref{eq.3}, is shown in Fig.~\ref{angle}(e). The topological edge states contribution ratio, defined as $\frac{\Delta \sigma(0)}{\sigma(0)}$ $\times$ 100\%, is calculated to be only $\sim$0.32\%, slightly smaller than that estimated from the field-dependent magnetoconductance data discussed above. This difference may stem from the anisotropic magnetoconductance of bulk states, which has not been considered previously. Such a small contribution of topological edge states is reasonable because topological edge states should only exist at the periphery of the topmost layers, as confirmed by the STM/STS measurements \cite{PRB-Feng}.
The same analysis was applied to the data for other fixed temperatures and the results are shown in Fig. S2 of the SM. With the temperature increasing, $\frac{\Delta \sigma(0)}{\sigma(0)}$ decreases to a negligible value due to the dephasing process. The first term of Eq.~\ref{eq.3} can nicely produce the experimental data above 5 K. The coherence length decreases from 16.2 to 13.1 nm as the temperature increases from 2 to 4 K and this monotonic reduction of coherence length also validates the applicability of this analysis.

\begin{figure}[]
\begin{center}
\includegraphics[width=0.46\textwidth]{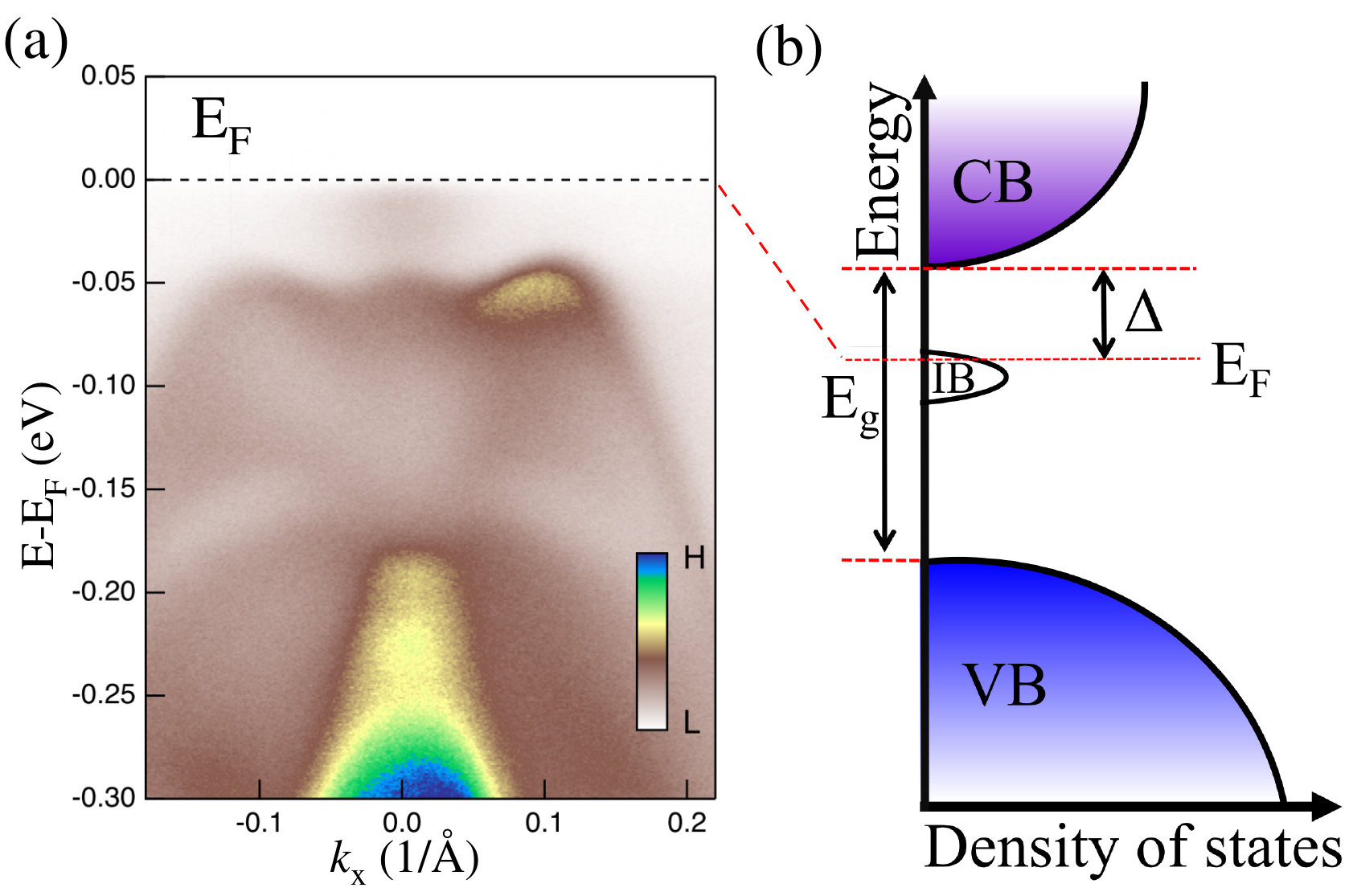}
\caption{\label{band} (a) ARPES band structure at 10 K along $\Gamma$-Y high symmetry direction corresponding to the intrachain
direction in real space. (b) Schematic picture of the energy diagram of the density of states of the bulk valence band (VB), bulk conduction band (CB) and impurity band (IB).}
\end{center}
\end{figure}

To obtain a better understanding of the transport properties of Ta$_2$Pd$_3$Te$_5$, we measured its band structure using ARPES.
Figure~\ref{band}(a) displays the band structure measured at 10 K along $\Gamma$-Y high symmetry direction corresponding to the intrachain
direction in real space, from which a spectral weight within the bulk band gap around the $\Gamma$ point can be observed clearly at the Fermi level. This spectral weight with negligible band dispersion is most possibly contributed by the impurity band (more supportive data  will be shown and discussed in details elsewhere), which is formed by the donators. Taking into consideration all the results discussed above, we can roughly conclude with a comprehensive picture about the transport mechanism in Ta$_2$Pd$_3$Te$_5$. The chemical potential is pinned in the impurity band at low temperatures. The impurity band contributes the finite electron-type bulk conduction with a carrier density 2.55 $\times$ 10$^{18}$ cm$^{-3}$ at 2.5 K as discussed above. The chemical potential is shifted slightly with the change of temperature, which is thus responsible for the temperature dependence of carrier density.
In the impurity band, both localized and extended electrons coexist, the boundary of which in the band structure is known as the mobility edge \cite{EPDS}. Hopping conduction of localized electrons takes place alongside the degenerate conduction of the impurity band with the temperature increasing, which gives rise to the VRH conduction as seen in Figs.~\ref{fig1}(d-e). At low temperatures, the degenerate bulk impurity band
also contributes to a metallic channel alongside the VRH conduction. As the temperature increasing higher, thermal activation of electrons from the donators to the bulk conduction bands dominates, giving rise to the semiconductor behavior and the increase of the electron-type charge carriers at high temperatures.
Schematic picture of the energy diagram of the density of states of the bulk and impurity bands is plotted in Fig.~\ref{band}(b), where $E_g$ denotes
the bulk energy gap between the bulk valence band and conduction band.
The WAL effect is mostly caused by the strong SOC of bulk states. In the meantime, we can find the tiny trace of the contribution from the topological
edge states from the anisotropic magnetoconductance exhibiting the WAL effect at low temperatures.

\section{Summary}

In summary, we have carried out comprehensive studies on the transport properties of the Ta$_2$Pd$_3$Te$_5$ single crystals.
The resistivity of Ta$_2$Pd$_3$Te$_5$, as one of TCLs with the existence of helical edge states, shows a semiconductor behavior from 300 K to 100 K, and crosses over to a three-dimensional VRH behavior until 18 K. At low temperatures, a large bulk resistivity exceeding 2 $\Omega$ cm can be found and tends to saturate, indicating a metallic contribution. The extremely small Hall mobility at low temperatures exclude the possibility of the dominance of topological edge states. The analysis of field-dependent and angle-dependent magnetoconductance suggests the exhibited WAL effect, originated from both topological edge states and bulk states. The WAL effect is revealed and analyzed quantitatively based on the two-dimensional HLN model. The coincidence of the fitting curves and the experimental data validate the perfect applicability of the HLN model. The contribution of the edge states in the total contribution may be as small as $\sim$0.32\%. Such a small portion hardly plays a role in the temperature-dependent resistivity and Hall effect, consistent with our data analysis. The quasi-1D Ta$_2$Pd$_3$Te$_5$, highly feasible by the mechanical exfoliation, provides another platform for investigations of the WAL effect and low-dimensional topological physics.

\begin{center}
{\bf ACKNOWLEDGEMENTS}
\end{center}

This work was supported by Zhejiang Provincial Natural Science Foundation of China (Grant No. LZ23A040002) and the National Natural Science Foundation of China (Grant No. 11504329). S.-Z.X. was supported by the National Natural Science Foundation of China (Grant No. 12204495). J.-K.B. was supported by the National Natural Science Foundation of China (Grant No. 12204298). X.-F.X. acknowledges support from the National Natural Science Foundation of China (Grants No. 12274369 and No. 11974061).

\bibliography{Jiao_Ta2Pd3Te5}

\end{document}